\documentclass[conference]{IEEEtran}
\IEEEoverridecommandlockouts
\usepackage[top=0.75in, bottom=0.75in, left=0.63in, right=0.63in]{geometry}

\usepackage{cite}
\usepackage{amsmath}
\usepackage{times}
\usepackage{epsfig}
\usepackage{graphicx}
\usepackage{subfig}
\usepackage{amsmath}
\usepackage{amssymb}
\usepackage{tabularray}
\usepackage{tikz}
\usepackage[ruled,vlined]{algorithm2e}  
\usepackage{stmaryrd}

\usepackage{dblfloatfix}
\usepackage{caption}  
\usepackage{longtable} 
\usepackage{url}
\usepackage{subcaption}
\usepackage{xurl}
\usepackage{afterpage}
\usepackage{comment}

\usepackage{makecell} 

\usepackage{booktabs} 
\usepackage{array}
\usepackage{longtable}
\usepackage[table]{xcolor}
\usepackage[utf8]{inputenc}
\usepackage[T1]{fontenc}
\usepackage{graphicx}
\usepackage{tabularx}
\usepackage{caption}
\usepackage{adjustbox}
\usepackage{supertabular} 
\usepackage{multirow}      
\usepackage{booktabs}      
\usepackage{lipsum}        
\usepackage{hyperref}  
\usepackage{url}       

\ifCLASSINFOpdf
\else
\fi

\begin{document}
%
\title{A Sensing Dataset Protocol for Benchmarking and Multi-Task Wireless Sensing}
%
%

%
%


\author{Jiawei Huang$^{*}$, Di Zhang$^{*}$, Yuanhao Cui, Xiaowen Cao, Tony Xiao Han, Xiaojun Jing, Christos Masouros
\thanks{Jiawei Huang$^{*}$, Di Zhang$^{*}$, Yuanhao Cui, and Xiaojun Jing are with the School of Information and Communication Engineering, Beijing University of Posts and Telecommunications, Beijing 100876, China. (e-mail: jiawei.aki, amandazhang, yuanhao.cui, jxiaojun@bupt.edu.cn). Huang and Zhang contributed equally to this manuscript.}

\thanks{Xiaowen Cao is with the College of Electronics and Information Engineering,
Shenzhen University, Shenzhen 518060, China (e-mail: caoxwen@szu.edu.cn).}

\thanks{Tony Xiao Han is with the Huawei Technology Co. Ltd., Guangdong 518129, China (e-mail: tony.hanxiao@huawei.com).}

\thanks{Christos Masouros is with the Department of Electronic and Electrical
Engineering, University College London, Torrington Place, London, WC1E
7JE, UK (email: c.masouros@ucl.ac.uk)}

\thanks{Yuanhao Cui is the corresponding author.}
}

\markboth{Journal of \LaTeX\ Class Files,~Vol.~14, No.~8, August~2015}%
{Shell \MakeLowercase{\textit{et al.}}: Bare Demo of IEEEtran.cls for IEEE Journals}

%



\maketitle

\begin{abstract}
Wireless sensing has become a fundamental enabler for intelligent environments, supporting applications such as human detection, activity recognition, localization, and vital-sign monitoring. Despite rapid advances, existing datasets and pipelines remain fragmented across sensing modalities, hindering fair comparison, transfer, and reproducibility. We propose the Sensing Dataset Protocol (SDP), a protocol-level specification and benchmark framework for large-scale wireless sensing. SDP defines how heterogeneous wireless signals are mapped into a unified perception data-block schema through lightweight synchronization, frequency–time alignment, and resampling, while a Canonical Polyadic-Alternating Least Squares (CP-ALS) pooling stage provides a task-agnostic representation that preserves multipath, spectral, and temporal structures. Built upon this protocol, a unified benchmark is established for detection, recognition, and vital-sign estimation with consistent preprocessing, training, and evaluation. Experiments under the cross-user split demonstrate that SDP significantly reduces variance ($\approx$88\%) across seeds while maintaining competitive accuracy and latency, confirming its value as a reproducible foundation for multi-modal and multi-task sensing research.

\end{abstract}


%
\IEEEpeerreviewmaketitle

\section{Introduction}\label{section1}
%
%
%
%

Wireless sensing has emerged as a key enabler for next-generation smart environments, supporting applications such as presence detection, person identification, activity recognition, and vital-sign monitoring. Compared with vision- or wearable-based systems, it leverages ubiquitous infrastructure, enables device-free operation, and remains robust under low-light or occluded conditions while alleviating privacy concerns. These advantages make it an indispensable component of integrated sensing and communications (ISAC), which is expected to play a crucial role in sixth-generation (6G) networks and beyond \cite{10908560,9737357,miao2025wi,10255711}. With higher frequencies, wider bandwidths, and large-scale antenna arrays, the wireless medium has evolved from a communication channel to a pervasive sensing resource, powering applications in healthcare, autonomy, and human–computer interaction \cite{10387517}.

To advance this field, several representative datasets have been developed. Widar3.0 supports Wi-Fi CSI-based gesture and activity recognition, SignFi targets device-free identification, M\textsuperscript{3}SC enables controlled multi-band and multi-weather simulations, and DeepSense 6G provides large-scale multimodal data \cite{zhang2021widar3,alkhateeb2023deepsense,10315088}. Algorithmic progress has also been notable: \cite{fu2024wi} used CNN–BiLSTM encoders for through-wall recognition, \cite{8514811} applied attention-based BLSTMs for cross-domain transfer, and \cite{strohmayer2024data} enhanced robustness with GAN-based augmentation and domain adaptation.

However, wireless sensing research still faces a fundamental limitation: data collection and expansion remain difficult.
CSI and other physical-layer measurements are strongly tied to device hardware parameters (e.g., RF chains, antenna spacing, calibration, bandwidth), resulting in heterogeneous data formats across platforms and frequency bands. Consequently, models trained on one device or frequency domain often fail to generalize to another. The absence of a standardized protocol for sensing data representation and preprocessing hinders cross-device integration, reproducibility, and fair benchmarking \cite{shi2022device,9729785}. Most prior works focused on dataset construction or algorithmic improvements, but not on establishing a common specification that defines how heterogeneous data should be represented, synchronized, and compared.

To address these challenges, this paper introduces the Sensing Dataset Protocol (SDP), a specification built upon the Sensing Dataset Platform—a large-scale (400+ hours, 14.1 TB) field-measured dataset initiated by IEEE ISAC–ETI\cite{sdp8} aimed at promoting reproducible wireless sensing research. Instead of standardizing low-level data formats, SDP defines a canonical tensor layout to unify heterogeneous measurements via lightweight synchronization, frequency-time alignment, and resampling, while preserving their physical semantics. Based on this protocol, we provide a reproducible end-to-end pipeline and a unified baseline model. This model utilizes a shared encoder and lightweight task-specific heads for diverse applications—including detection, recognition, and vital-sign estimation—thus establishing a scalable and reproducible benchmark for the wireless sensing community.



The remainder of this paper is organized as follows. Section \ref{section2} formalizes the SDP data-block schema and mapping procedure. Section \ref{section3} describes the unified model design and training/inference protocol built upon SDP. Section \ref{section4} details the experimental setup and reports results under fair splits. Section \ref{section5} concludes the paper and discusses future extensions.

\begin{figure*}[thbp!]
    \centering
    \captionsetup{justification=raggedright}
    \includegraphics[width=0.8\linewidth]{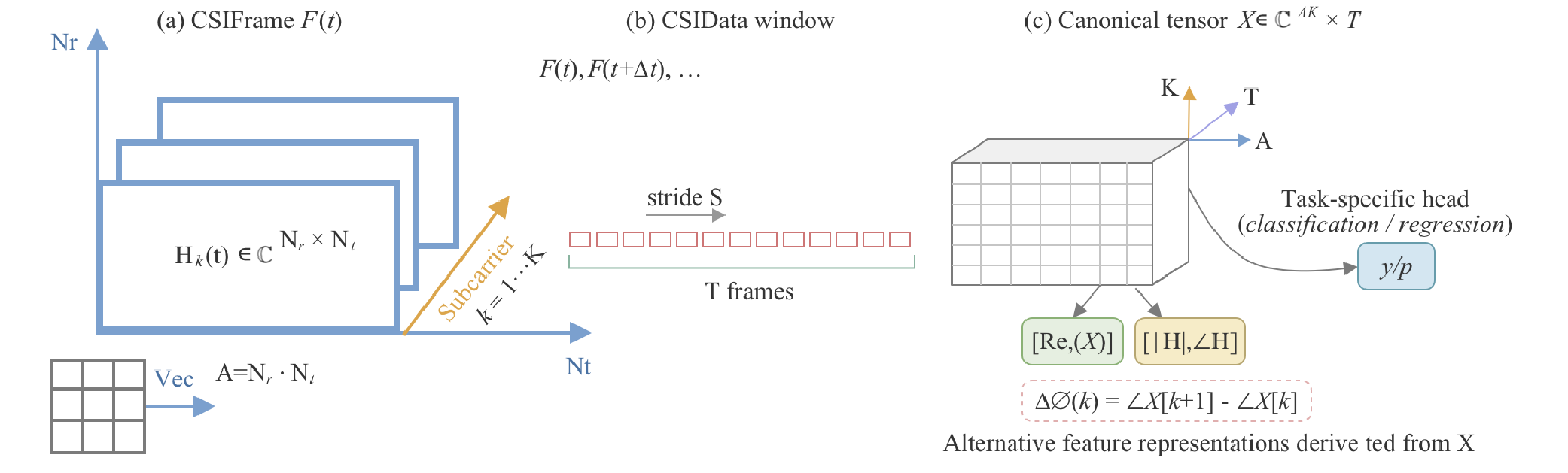}
    \caption{Protocol-defined data organization in SDP. Each packet-level CSIFrame contains timestamped per-subcarrier matrices, which are aggregated through a sliding window into CSIData samples and tensorized into the canonical (A, K, T) layout used by all tasks.
}
    \label{fig:sdp_csi_tensor}
\end{figure*}

\section{System Model}\label{section2}

\subsection{Overall System Model}

This work uses channel state information (CSI) as the principal example of wireless sensing data to illustrate how the protocol operates. CSI captures the frequency–domain channel response between each transmit–receive antenna pair across multiple subcarriers and over time, encoding both static multipath components and dynamic perturbations induced by moving objects or human activities. Compared with coarse indicators such as received signal strength indicator (RSSI), CSI provides richer spatio–temporal structure and has been widely used in Wi-Fi and cellular sensing studies~\cite{shi2022device,9729785}.

Consider an orthogonal frequency-division multiplexing (OFDM) system with $N_t$ transmit and $N_r$ receive antennas. The CSI of the $k$-th subcarrier at time $t$ is
\begin{equation}
    \mathbf{H}_k(t)\in\mathbb{C}^{N_r\times N_t},
\end{equation}
where $h_{r,t}^{(k)}(t)$ denotes the complex gain between transmit antenna $t$ and receive antenna $r$. In a multipath environment with $L$ paths,
\begin{equation}
    h_{r,t}^{(k)}(t)=\sum_{\ell=1}^{L}\alpha_\ell\,e^{-j2\pi f_k\tau_\ell(t)}\,e^{-j\phi_\ell(t)},
\end{equation}
where $\alpha_\ell$ is the complex attenuation of path $\ell$, $\tau_\ell(t)$ is the delay, $f_k$ is the subcarrier frequency, and $\phi_\ell(t)$ models phase rotation due to device impairments and carrier-frequency offset. Static reflectors (e.g., walls) yield time-invariant terms, whereas dynamic reflectors (e.g., human motion) induce time variation.

Across consecutive OFDM packets, per-packet CSI forms a third-order tensor
\begin{equation}
    \mathcal{H}(t)\in\mathbb{C}^{N_r\times N_t\times K},
\end{equation}
with $K$ subcarriers. Over a window of $T$ packets,
\begin{equation}
    \mathbf{X}\in\mathbb{C}^{N_r\times N_t\times K\times T},
\end{equation}
which preserves spatial diversity (antennas), spectral diversity (subcarriers), and temporal evolution. Measurements are corrupted by noise and hardware imperfections; additive white Gaussian noise (AWGN) is modeled as
\begin{equation}
    \tilde{h}_{r,t}^{(k)}(t)=h_{r,t}^{(k)}(t)+n_{r,t}^{(k)}(t),\qquad n_{r,t}^{(k)}(t)\sim\mathcal{CN}(0,\sigma^2).
\end{equation}
Packet-level asynchrony, common phase error, and sampling-frequency offset introduce further distortions. Typical remedies include phase sanitization, adjacent-subcarrier phase-difference construction, and amplitude normalization. While OFDM is used for exposition, the protocol discussed next targets a unified learning interface that heterogeneous sensing waveforms can map to with lightweight synchronization and frequency–time resampling.

\subsection{Sensing Data-Block Schema under the SDP}

Guided by Fig.~\ref{fig:sdp_csi_tensor}, the SDP defines a unified sensing data-block schema and mapping interface to provide reproducible, semantically consistent inputs for learning models while preserving essential physical semantics.

At the frame level, each CSIFrame at time $t$ contains a timestamp and a set of per-subcarrier channel matrices $\{\mathbf{H}_k(t)\}_{k\in\mathcal{K}}$ with a documented antenna-pair order. Frames are strictly time-ordered. At the sequence level, consecutive frames are aggregated using a sliding window of length $T$ and stride $S$, forming a canonical tensor whose axes correspond to antenna pairs ($A\!=\!N_rN_t$), subcarriers ($K$), and temporal steps ($T$). Optional padding is recorded by a binary validity mask, and timestamps remain unchanged.

To feed real-valued models without discarding phase, two parameterizations are supported: real–imaginary stacking and amplitude–phase stacking. An adjacent-subcarrier phase-difference branch can be enabled to improve robustness to common phase error and sampling offsets. Unless otherwise stated, amplitudes are converted to dB and z-score normalized per session, while subcarrier indices and antenna-pair ordering are fixed across all splits.

Heterogeneous capture tools and chipsets are mapped to the canonical $(A,K,T)$ space through declarative adapters consistent with Fig.~\ref{fig:sdp_csi_tensor}; per-band subcarrier lists and antenna mappings are documented and versioned. This protocol allows heterogeneous sensing data to be projected into a unified interface for preprocessing, training, and evaluation without imposing a rigid CSI standard. While the schema is described per device for clarity, it inherently supports multi-device scenarios such as Widar, where multiple receivers observe the same physical event from different viewpoints. This design enables correlated-data modeling and lays the groundwork for future collaborative and distributed sensing within the SDP framework.

\section{Proposed Learning-Based Method}
\label{section3}

Designed for scalability, the learning framework operates uniformly across large datasets and diverse sensing modalities, supporting consistent training under constrained tuning budgets. Building upon the unified sensing data-block representation defined by the SDP in Section~\ref{section2}, this section formulates the learning framework that operates uniformly across sensing tasks.

\subsection{Problem Formulation}
All tasks operate on a canonical windowed tensor $\mathbf{X}\in\mathbb{C}^{A\times K\times T}$.
For real-valued networks, we adopt real–imag stacking to obtain
$\tilde{\mathbf{X}}\in\mathbb{R}^{A\times K\times T\times 2}$, reshaped to $(B,T,K,A\times 2)$ before the encoder.

Heterogeneous captures are projected into this canonical layout using a fixed window length $W$ and stride $S$, followed by session-wise normalization, phase sanitization, deterministic reindexing, and light denoising. 
Labels are aligned to the window center timestamp.

A task-agnostic representation is then obtained via rank-$R$ CP decomposition estimated by Alternating Least Squares (ALS), which factorizes the tensor as
\begin{equation}
\label{eq:cp_factorization}
\mathbf{X} \approx \sum_{r=1}^{R} \mathbf{a}_r \circ \mathbf{b}_r \circ \mathbf{c}_r
= [\![\mathbf{A},\mathbf{B},\mathbf{C}]\!],
\end{equation}
producing factor matrices $(\mathbf{A},\mathbf{B},\mathbf{C})$ and a deterministic pooled descriptor $\mathbf{h}$.
This operation extracts separable spatial, spectral, and temporal patterns while preserving the multipath structure.

Each update alternately minimizes the reconstruction error along one tensor mode while keeping the others fixed, yielding a lightweight unsupervised feature extractor. 
ALS updates solve regularized normal equations on mode unfoldings:
\begin{subequations}
\label{eq:als_updates}
\begin{align}
\mathbf{A}&\leftarrow \mathbf{X}_{(1)}(\mathbf{C}\odot\mathbf{B})
\big[(\mathbf{C}\odot\mathbf{B})^{\top}(\mathbf{C}\odot\mathbf{B})+\epsilon\mathbf{I}\big]^{-1},\\
\mathbf{B}&\leftarrow \mathbf{X}_{(2)}(\mathbf{C}\odot\mathbf{A})
\big[(\mathbf{C}\odot\mathbf{A})^{\top}(\mathbf{C}\odot\mathbf{A})+\epsilon\mathbf{I}\big]^{-1},\\
\mathbf{C}&\leftarrow \mathbf{X}_{(3)}(\mathbf{B}\odot\mathbf{A})
\big[(\mathbf{B}\odot\mathbf{A})^{\top}(\mathbf{B}\odot\mathbf{A})+\epsilon\mathbf{I}\big]^{-1}.
\end{align}
\end{subequations}

Here $\odot$ denotes the Khatri–Rao product and $\mathbf{X}_{(n)}$ is the mode-$n$ unfolding.

This formulation aligns with the separable structure of multipath propagation across antennas, subcarriers, and time. 
Under mild diversity, identifiability up to scaling and permutation holds when
\begin{equation}
\label{eq:kruskal}
k_{\mathbf A}+k_{\mathbf B}+k_{\mathbf C}\ \ge\ 2R+2,
\end{equation}
where $k_{\mathbf A}$ denotes the Kruskal rank (krank) of $\mathbf A$. 
Each ALS sweep involves matrix–tensor products and small linear systems, leading to a cost that scales linearly with tensor size and rank, and thus a lightweight, label-free extractor. 
A deterministic pooling operator then aggregates per-component statistics (weights, spatial concentration, temporal/spectral centroids and spreads) together with global summaries into a fixed-length feature vector $\mathbf{h}$ shared by all tasks.

Given the task set
$\mathcal{T} $= \{\texttt{detection},\, \texttt{recognition},\, \texttt{vital-sign estimation}\},
each task consumes the same $\mathbf{h}$ and differs only by a lightweight head (logistic/softmax classifiers for detection/recognition and a shallow regressor for vital signs). 
The empirical risk is minimized as
\begin{equation}
\label{eq:objective}
\begin{aligned}
\min_{\{\theta_\tau\}}\ \mathcal{L}
&= \frac{1}{N}\sum_{i=1}^{N}\sum_{\tau\in\mathcal{T}}
\lambda_{\tau}\,\ell_{\tau}\!\big(g_{\tau}(\mathbf{h}^{(i)};\theta_{\tau}),\,\mathbf{y}^{(i)}_{\tau}\big) \\
&\quad + \lambda_{\mathrm{reg}}\sum_{\tau\in\mathcal{T}}\|\theta_{\tau}\|_2^2,
\end{aligned}
\end{equation}
where $\lambda_{\tau}\!\ge\!0$ are task weights (default $1$) and $\lambda_{\mathrm{reg}}$ controls weight decay. 
Unless otherwise noted, $(W,S)$, the canonical ordering of $(A,K,T)$, all preprocessing steps, the CP rank $R$, and the pooling recipe are fixed to ensure fair comparisons across datasets and tasks.

\begin{figure*}[thtp!]
    \centering
    \includegraphics[width=0.8\linewidth]{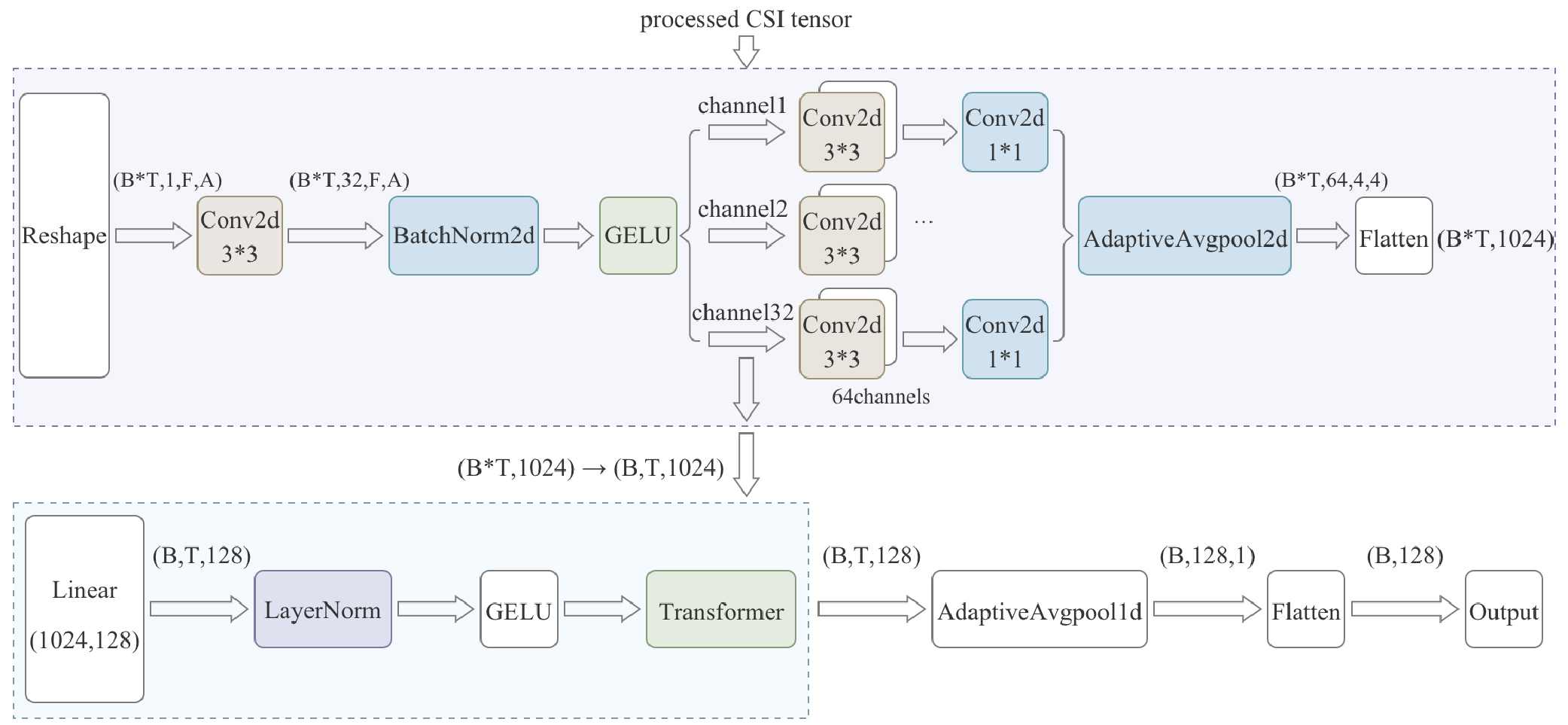}
    \captionsetup{justification=raggedright}
    \caption{Network architecture of the SDP benchmark model. Each window of CSI data is reshaped and processed by 2D spatial convolutions and a temporal Transformer module to generate task-specific features for detection, recognition, and vital-sign estimation. Dimensions are shown for a representative configuration.}
    \label{fig:deeplearning_based_architecture}
\end{figure*}

\subsection{Benchmark Model Design}

Building upon the unified sensing data-block schema defined by the SDP, a benchmark model is constructed to map the protocol-compliant representation to multiple task outputs through a single, reproducible training and inference pipeline. 
The goal is to provide a low-variance, comparable, and extensible baseline across heterogeneous datasets and sensing tasks.

\begin{algorithm}[!bt]
\caption{Unified multi-task training under uncertainty-based loss weighting.}
\label{alg:train}
\KwIn{Mini-batches $\{(\mathbf{X},\{\mathbf{y}_\tau\}_{\tau\in\mathcal{T}})\}$; shared encoder $\phi$; task heads $\{g_\tau\}$}
\KwOut{Parameters $\theta=\{\theta_{\mathrm{enc}},\theta_\tau\}$ and uncertainties $\{\sigma_\tau\}$}
Initialize $\theta$; initialize $\{\sigma_\tau>0\}$\;
\For{epoch $=1,\dots,E$}{
    Sample a mini-batch and compute $\mathbf{z}\leftarrow \phi(\mathbf{X};\theta_{\mathrm{enc}})$\;
    \For{each $\tau\in\mathcal{T}$}{
        $\hat{\mathbf{y}}_\tau \leftarrow g_\tau(\mathbf{z};\theta_\tau)$; compute task loss $\ell_\tau$\;
    }
    \If{domain alignment enabled}{
        compute $\ell_{\mathrm{align}}(\mathbf{z})$;
    }
    $\mathcal{L}\leftarrow \sum_{\tau}\frac{1}{2\sigma_\tau^2}\ell_\tau
    +\sum_{\tau}\log\sigma_\tau
    +\lambda_{\mathrm{align}}\ell_{\mathrm{align}}$\;
    Update $\theta,\{\sigma_\tau\}$ with AdamW on $\nabla\mathcal{L}$\;
}
\end{algorithm}

\subsubsection{Input representation via CP--ALS pooling}
Each sensing window $\mathbf{X}$ from the SDP schema is factorized by CP--ALS as in \eqref{eq:cp_factorization} with updates in \eqref{eq:als_updates}. 
After each sweep, factor columns are $\ell_2$-normalized, their scales absorbed into nonnegative weights, and components sorted by descending weight. 
The pooled descriptor $\mathbf{h}$ is a fixed-length, deterministic summary used by all tasks. 
For deployments that benefit from explicit mode structure, $\mathbf{h}$ can be linearly reprojected to a compact $(\tilde A,\tilde K,\tilde T)$ tensor before entering the encoder. 
The rank $R$ and pooling recipe are fixed across datasets for comparability. 
This CP--ALS step serves as the protocol-defined bridge that transforms heterogeneous signal tensors into a consistent, device-agnostic representation without losing essential multipath, temporal, or spectral information.

\subsubsection{Shared encoder and lightweight task heads}
A thin shared encoder $\phi(\cdot)$ operates on $\mathbf{h}$ (or its reprojection) to extract joint spatio--spectral--temporal features. 
The encoder adopts separable attention across channel, frequency, and time axes followed by pointwise mixing; stacking $L$ residual blocks with layer normalization yields a compact backbone. 
Task heads $g_{\tau}(\cdot)$ are deliberately minimal: a logistic/softmax classifier for detection and recognition, and a shallow regressor for vital-sign estimation. 
The overall predictor is
\begin{equation}
    f_{\tau}=g_{\tau}\circ\phi,
\end{equation}
enforcing a single input schema and shared feature extractor across tasks, with task-specific differences confined to the head parameters. 
This structure ensures that performance variations primarily reflect sensing diversity or task difficulty rather than architectural mismatch.

\subsubsection{Network Architecture}
\label{sec:net-arch}
Building on the shared encoder and task heads, we instantiate a lightweight architecture with explicit spatial–temporal decomposition, as illustrated in Fig.~\ref{fig:deeplearning_based_architecture}. 
Given a processed CSI tensor $\tilde{\mathbf{X}}\!\in\!\mathbb{R}^{B\times T\times F\times A}$, each time slice is reshaped to $(B\!\cdot\!T,1,F,A)$ and fed into a 2D convolutional stem. 
The stem begins with a $3{\times}3$ Conv2d layer (1$\!\to$32 channels), followed by BatchNorm2d and a GELU activation to capture local coupling among adjacent subcarriers and antenna elements. 
Each of the resulting 32 feature channels is further expanded by a $3{\times}3$ Conv2d layer to 64 channels, followed by a $1{\times}1$ Conv2d for pointwise refinement. 
These operations preserve fine-grained frequency–antenna correlations while maintaining computational efficiency. 
An AdaptiveAvgPool2d layer then reduces the spatial map to $(4{\times}4)$, and the flattened output yields a $(B\!\cdot\!T,1024)$ descriptor, which is reshaped back to $(B,T,1024)$ to restore temporal order.

Subsequently, a linear bottleneck projects each step from 1024 to 128 dimensions, producing $(B,T,128)$ features. 
Layer normalization and a GELU activation are applied before feeding the sequence into a shallow Transformer block that models temporal dependencies. 
The Transformer consists of two layers with four attention heads, model width 128, feed-forward dimension 256, and dropout 0.2. 
The temporal output $(B,T,128)$ is adaptively averaged along the time dimension using AdaptiveAvgPool1d, resulting in a compact $(B,128)$ representation. 
Finally, this feature vector is flattened and passed to the task-specific output head $g_{\tau}(\cdot)$ for classification or regression, depending on the sensing task.

With a representative input $(T,F,A)=(64,56,3)$, the network contains approximately 4.3\,M parameters and requires 0.42\,GFLOPs per window. 
When evaluated on an NVIDIA A100 (40\,GB) GPU with batch size 1 and AMP enabled, the median single-window latency is 7.4\,ms (p90: 12.8\,ms). 
This design cleanly separates instantaneous spatial encoding from temporal modeling: the small $3{\times}3$ kernels efficiently capture local spatial context, the 1024$\!\to\!$128 linear bottleneck limits memory and FLOPs, and the short Transformer stack ensures scalable temporal reasoning with minimal overhead.

\subsubsection{Unified Training Protocol and Inference Pipeline}
Training follows the multi-task objective in \eqref{eq:objective} with standardized losses per task. 
Homoscedastic uncertainties $\{\sigma_\tau\}$ are learned to adaptively weight task losses. 
For cross-dataset or cross-device generalization, a domain-alignment regularizer (MMD or contrastive/InfoNCE) is optionally applied to the shared representation without modifying the input schema or CP--ALS pooling. 
Optimization uses AdamW with weight decay and cosine scheduling, together with early stopping on a validation split.

During inference, each window undergoes the same preprocessing and CP--ALS pooling to produce $\mathbf{h}$, which is then passed through the shared encoder and the task-specific head. 
The overall latency is primarily determined by the ALS sweeps and the encoder forward passes; both scale roughly linearly with tensor size and rank. 
To ensure fair comparison across datasets and tasks, $(W,S)$, the $(A,K,T)$ ordering, preprocessing steps, CP rank $R$, and pooling recipe remain fixed unless explicitly stated.

This benchmark model, defined on top of the SDP protocol, provides a reproducible and fair reference for evaluating sensing algorithms under diverse data sources and task configurations. 
It unifies preprocessing, factorization, feature encoding, and inference, ensuring that performance differences reflect genuine algorithmic advances rather than dataset-specific biases.

\begin{figure}[thtp!]
    \centering
    \setlength{\abovecaptionskip}{0.cm}
    \includegraphics[width=0.8\linewidth]{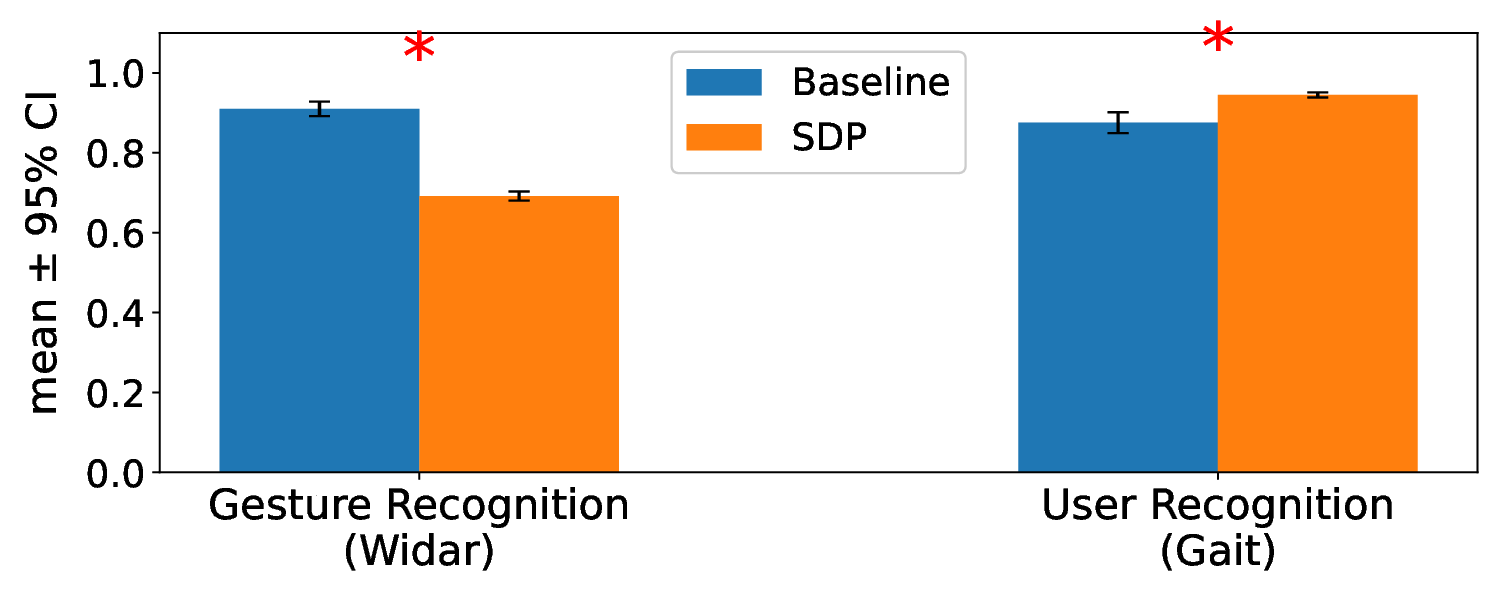}
    \captionsetup{justification=raggedright}
    \caption{Comparison of two recognition tasks. Bars show mean performance over 500 runs (five fixed seeds, 100 repeats each); 
error bars indicate 95\% confidence intervals (Student-$t$). On Gait, the SDP model significantly outperforms the baseline ($p<0.05$). 
Across both tasks, the SDP model exhibits narrower intervals, indicating improved stability.}

    \label{fig:res_bar}
\end{figure}

\begin{figure}[thp!]
    \centering
    \captionsetup{justification=raggedright}
    \subfloat[Average normalized confusion matrix on Widar]{
        \label{fig:conf_mat_widar}
        \includegraphics[width=0.48\linewidth]{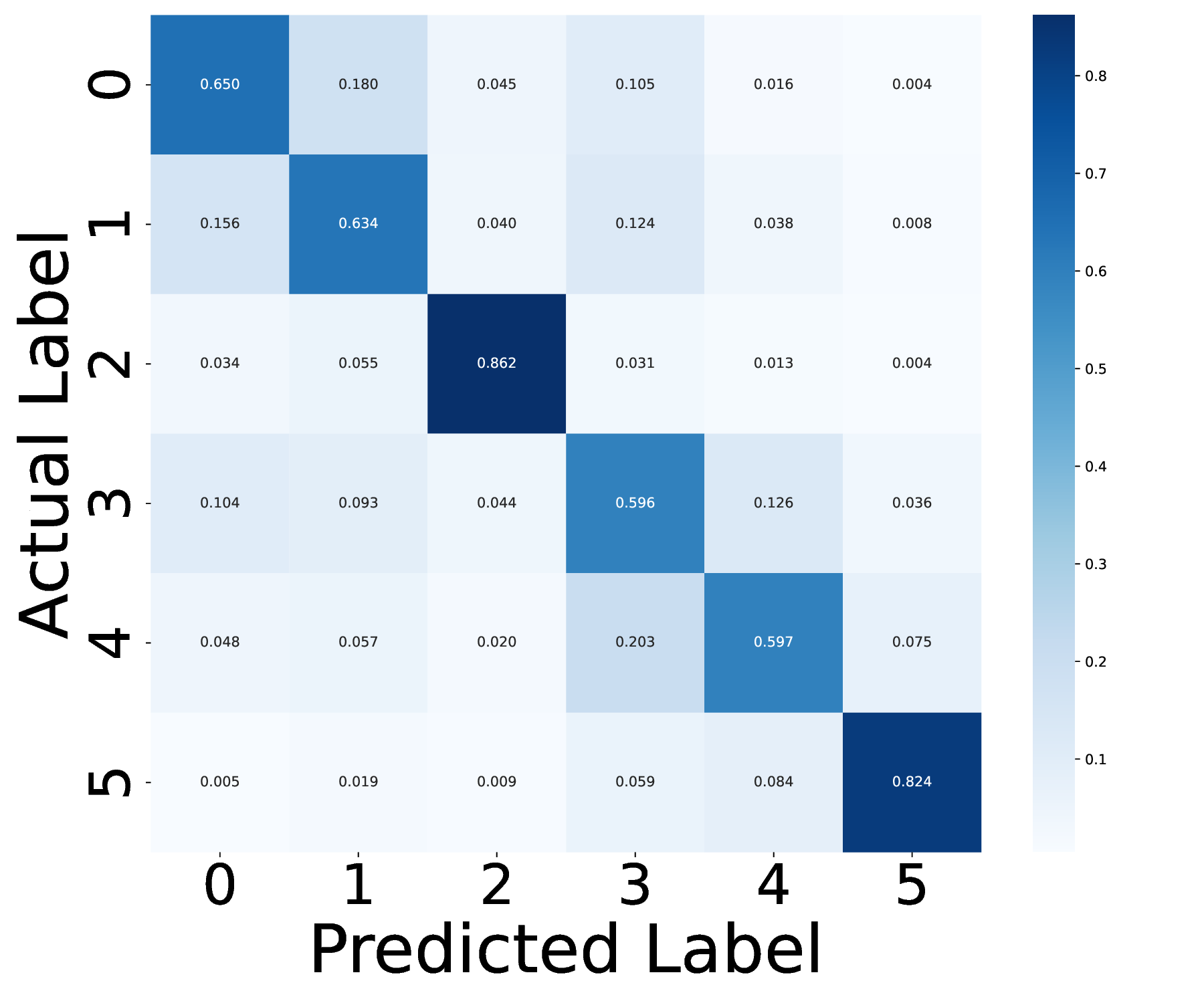}}
    \subfloat[Average normalized confusion matrix on Gait ]{
        \label{fig:conf_mat_gait}
        \includegraphics[width=0.48\linewidth]{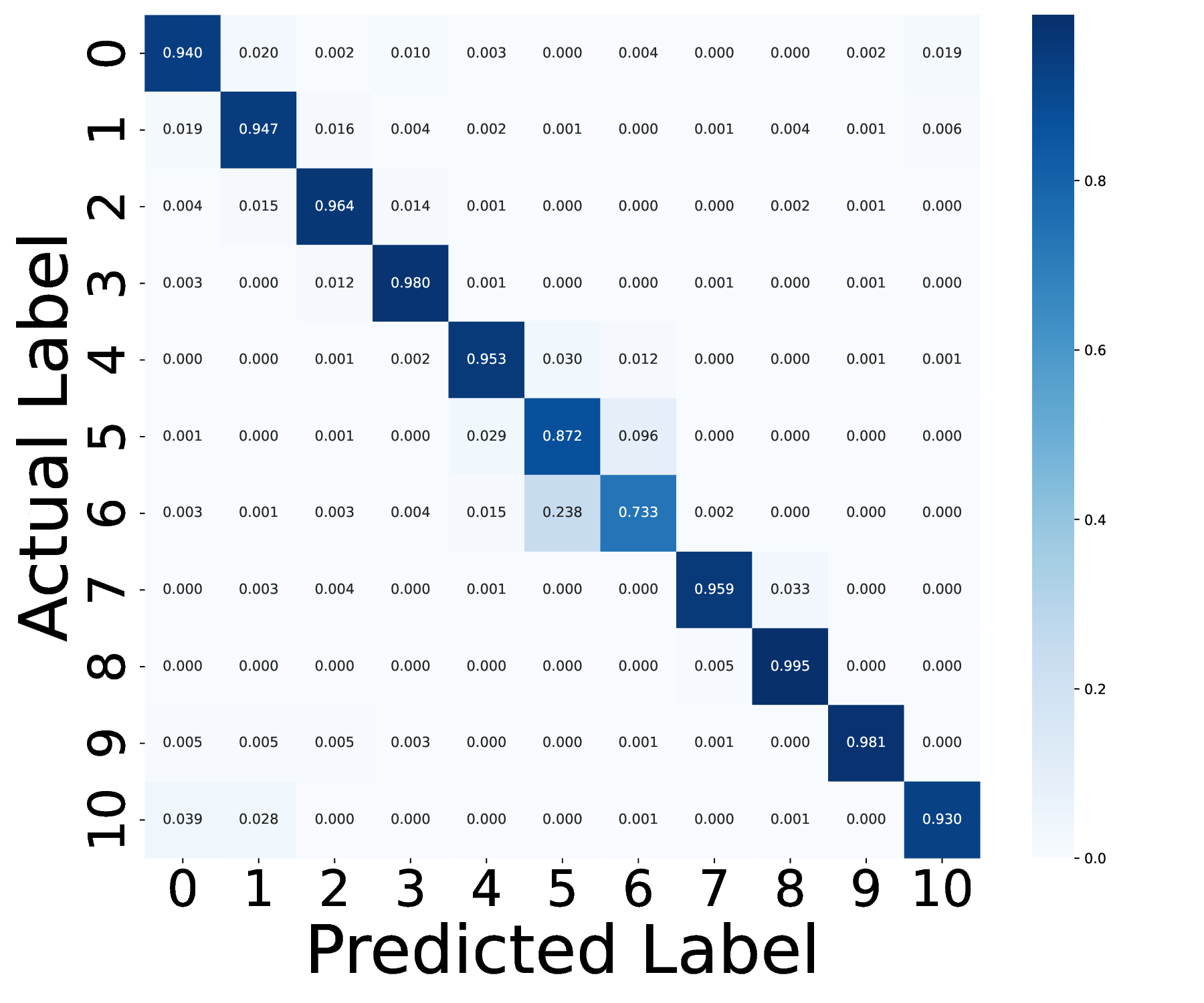}}
    \caption{Row-normalized confusion matrices averaged over 500 runs.}
    \label{fig:confusion_matrices}
\end{figure}

\section{Experiments and Results}
\label{section4}

\subsection{Experimental Setup}

All experiments adopt the \textit{cross-user} split to avoid inter-session leakage and ensure fair comparison. Unless otherwise stated, evaluation is performed at the window level, and clip-level aggregation is additionally reported for reference. A single unified configuration is used across datasets and tasks under the limited-tuning principle: fixed window/stride $(W,S)$ in the SDP protocol, CP rank $R$ with a fixed number of ALS sweeps, and AdamW with a cosine schedule and early stopping. Five fixed seeds $\mathcal{S}=\{992,863,702,443,542\}$ are used; each seed is repeated 100 times. Metrics are reported as mean$\pm$std across runs, and 95\% confidence intervals are computed with the Student-$t$ distribution. Classification tasks are evaluated by Top-1 accuracy and macro-F1 when applicable. Efficiency is assessed by parameter count, FLOPs, peak memory, and single-window inference latency (p50/p90, batch=1, post warm-up). All experiments run on a single NVIDIA A100 (40\,GB) with CUDA/cuDNN~12.6 and PyTorch~2.7.1. All dataset schemas, preprocessing scripts, and benchmark baselines will be publicly released at \url{https://sdp8.net} to support reproducibility and community benchmarking.

\subsection{Experimental Result}

Figure~\ref{fig:res_bar} summarizes the overall performance across the Widar and Gait datasets. 
The SDP model achieves 69.2$\pm$1.2\% Top-1 accuracy, whereas the native baseline attains 90.7$\pm$3.8\%. 
Although the mean accuracy is moderately lower, the dispersion across random seeds is considerably reduced: 
the variance drop $\mathrm{VarDrop}=1-\frac{s_{\mathrm{SDP}}^2}{s_{\mathrm{base}}^2}=88\%$ is statistically significant under the Brown–Forsythe test ($p_{\mathrm{var}}<0.01$). 
Bars in Fig.~\ref{fig:res_bar} report the mean over 500 independent runs (five fixed seeds, 100 repeats each) with 95\% confidence intervals computed using the Student-$t$ distribution; 
asterisks denote paired $t$-test significance ($p<0.05$). 
Overall, the SDP model yields narrower confidence bounds and more stable results across both tasks, confirming the protocol’s effectiveness in controlling random variability.

\begin{figure}[htbp]
    \centering
     \captionsetup{justification=raggedright}
    \subfloat[Element-wise std of the normalized confusion matrix on Widar]{
        \label{fig:stability_widar}
        \includegraphics[width=0.48\linewidth]{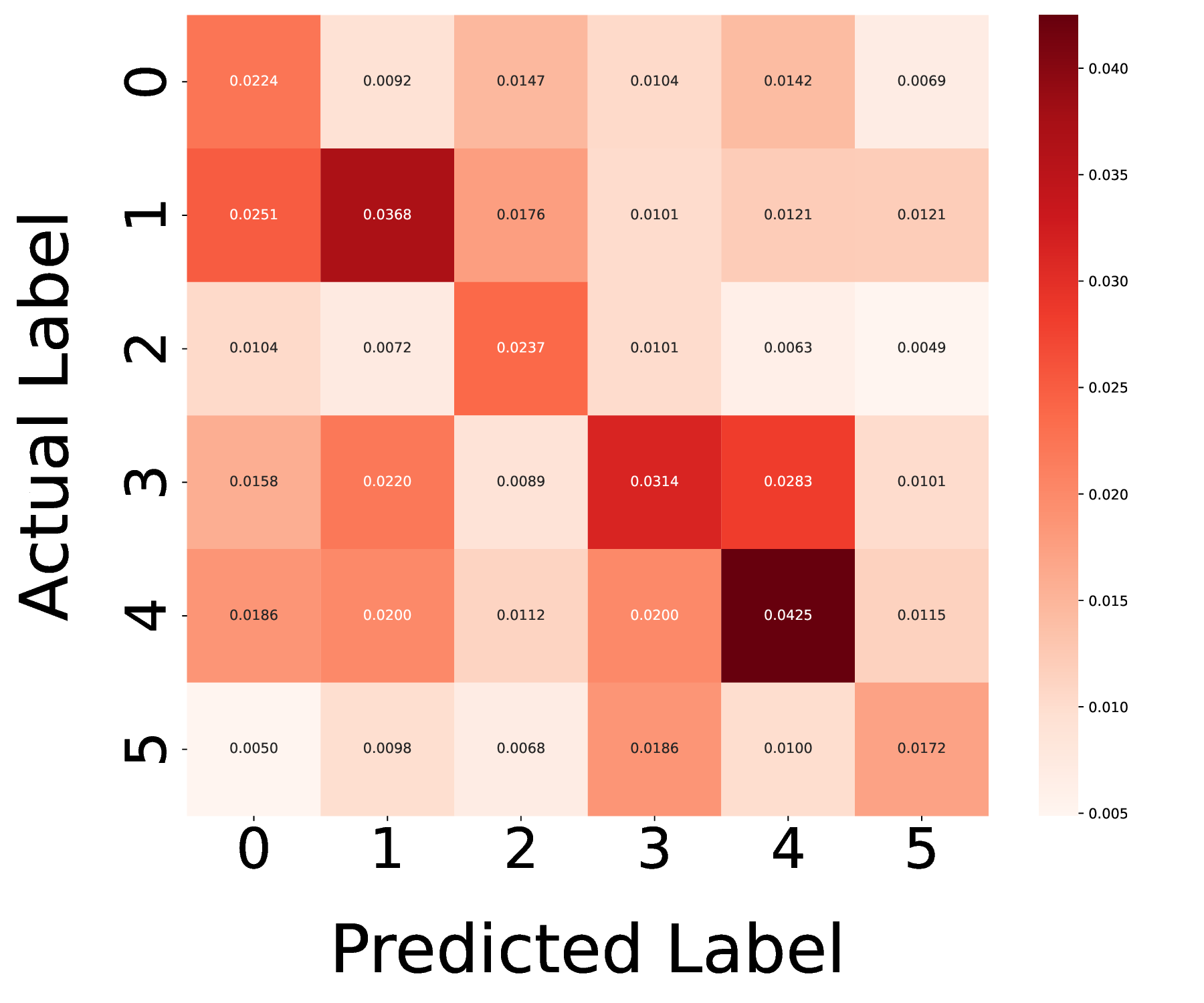}}
    \subfloat[Element-wise std of the normalized confusion matrix on Gait]{
        \label{fig:stability_gait}
        \includegraphics[width=0.48\linewidth]{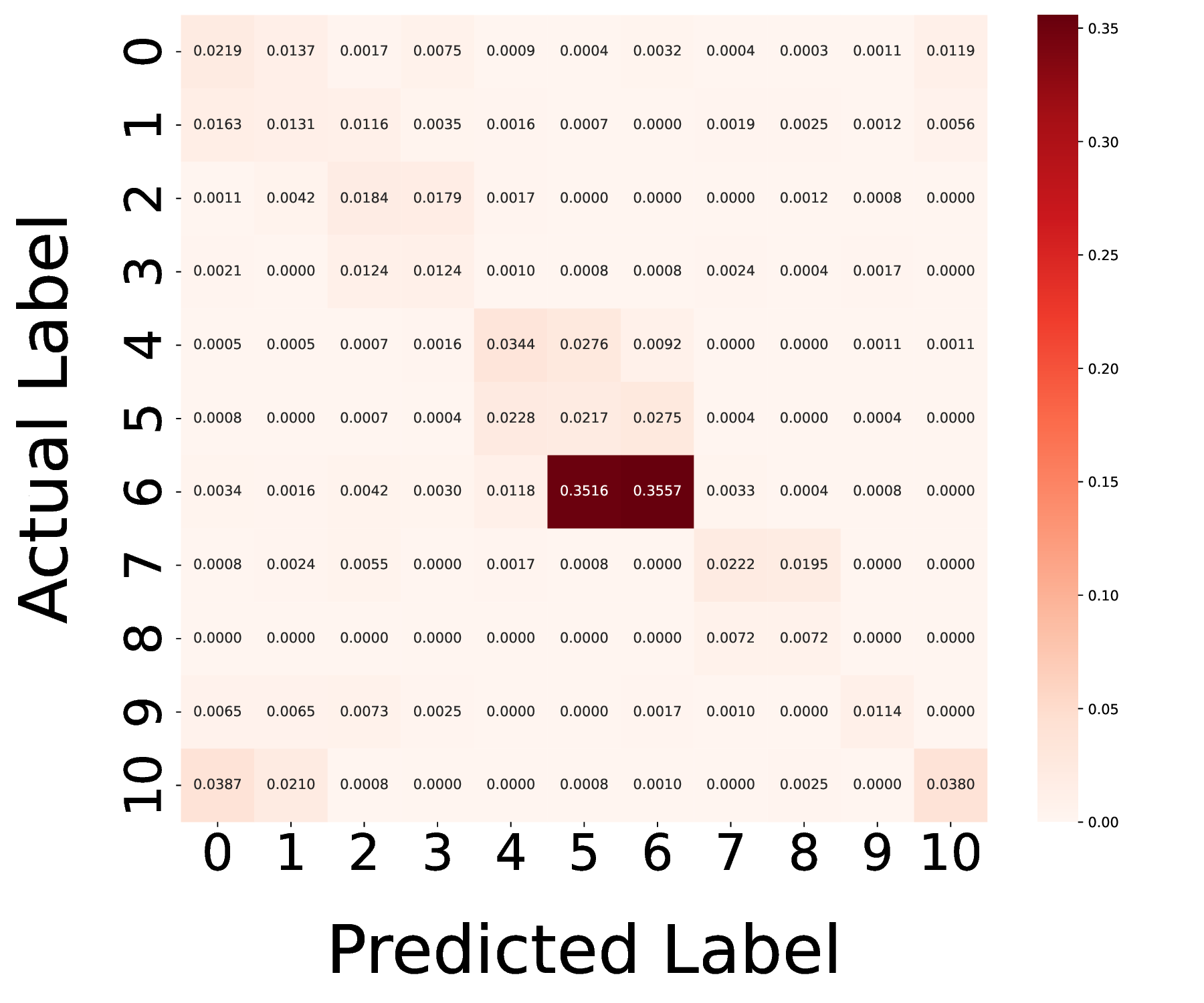}}
    \caption{Across-seed variability maps over 500 runs. Variance concentrates on sparse off-diagonal cells that match residual confusions in Fig.~\ref{fig:confusion_matrices}; diagonals remain low, indicating stable per-class recall.}
    \label{fig:stability_matrices}
\end{figure}

\begin{figure}[thp!]
    \centering
    \captionsetup{justification=raggedright}
    \setlength{\abovecaptionskip}{0.cm}
    \includegraphics[width=0.95\linewidth]{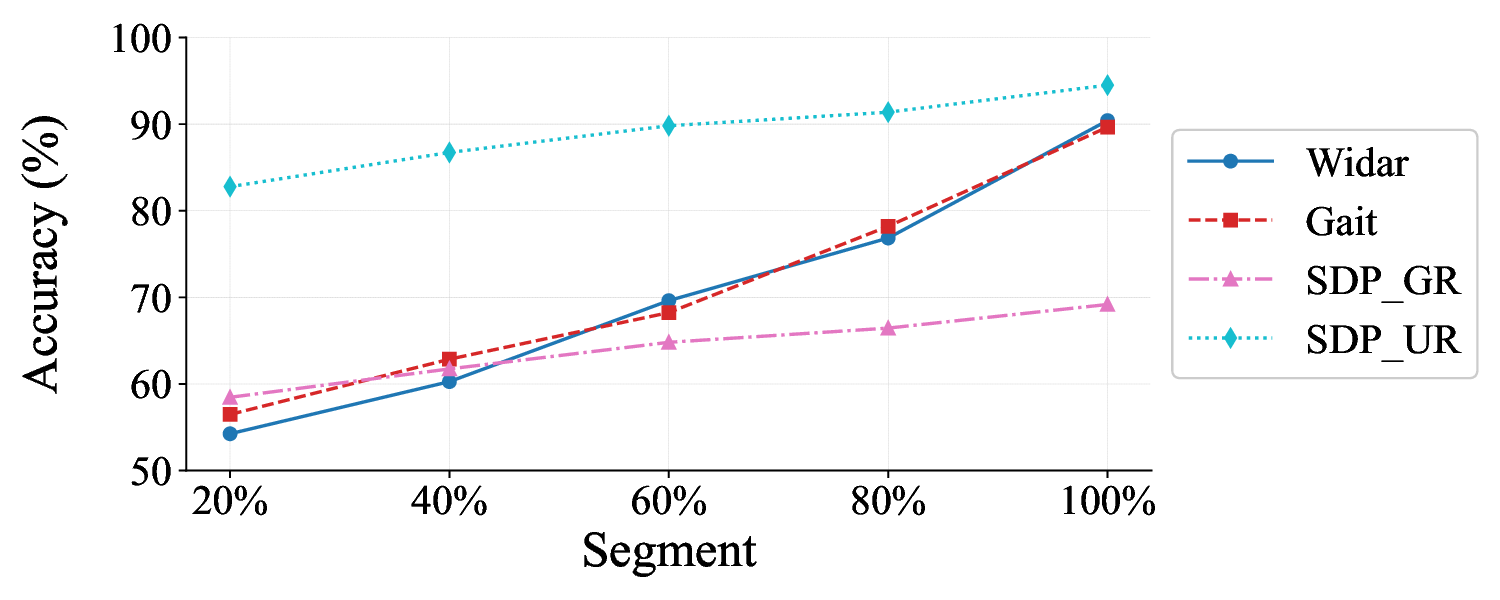}
    \caption{Comparison of model average inference accuracy versus the percentage of training data. GR for Gesture Recognition, UR for User Recognition.}
    \label{fig:res_curve}
\end{figure}

To visualize error consistency, Fig.~\ref{fig:confusion_matrices} shows the row-normalized confusion matrices averaged over five runs, 
while Fig.~\ref{fig:stability_matrices} presents their element-wise standard-deviation maps. 
Both perspectives consistently exhibit strong diagonals and limited, localized off-diagonal variance, demonstrating that misclassification patterns are systematic rather than random. 
This complementary analysis further supports the variance-reduction trend observed in Fig.~\ref{fig:res_bar}.

Label efficiency is assessed by subsampling the training data to \{20, 40, 60, 80, 100\}\% without retuning. The learning curves in Fig.~\ref{fig:res_curve} increase smoothly with the label ratio and show visibly narrower seed bands for the SDP model; using the 100\% native baseline as reference, the SDP model reaches 90.3\% of that accuracy with only 60\% labeled data, consistent with the regularizing effect of protocol-defined preprocessing and the fixed pooling/encoder.

\begin{figure}[htbp]
    \centering
    \includegraphics[width=0.8\linewidth]{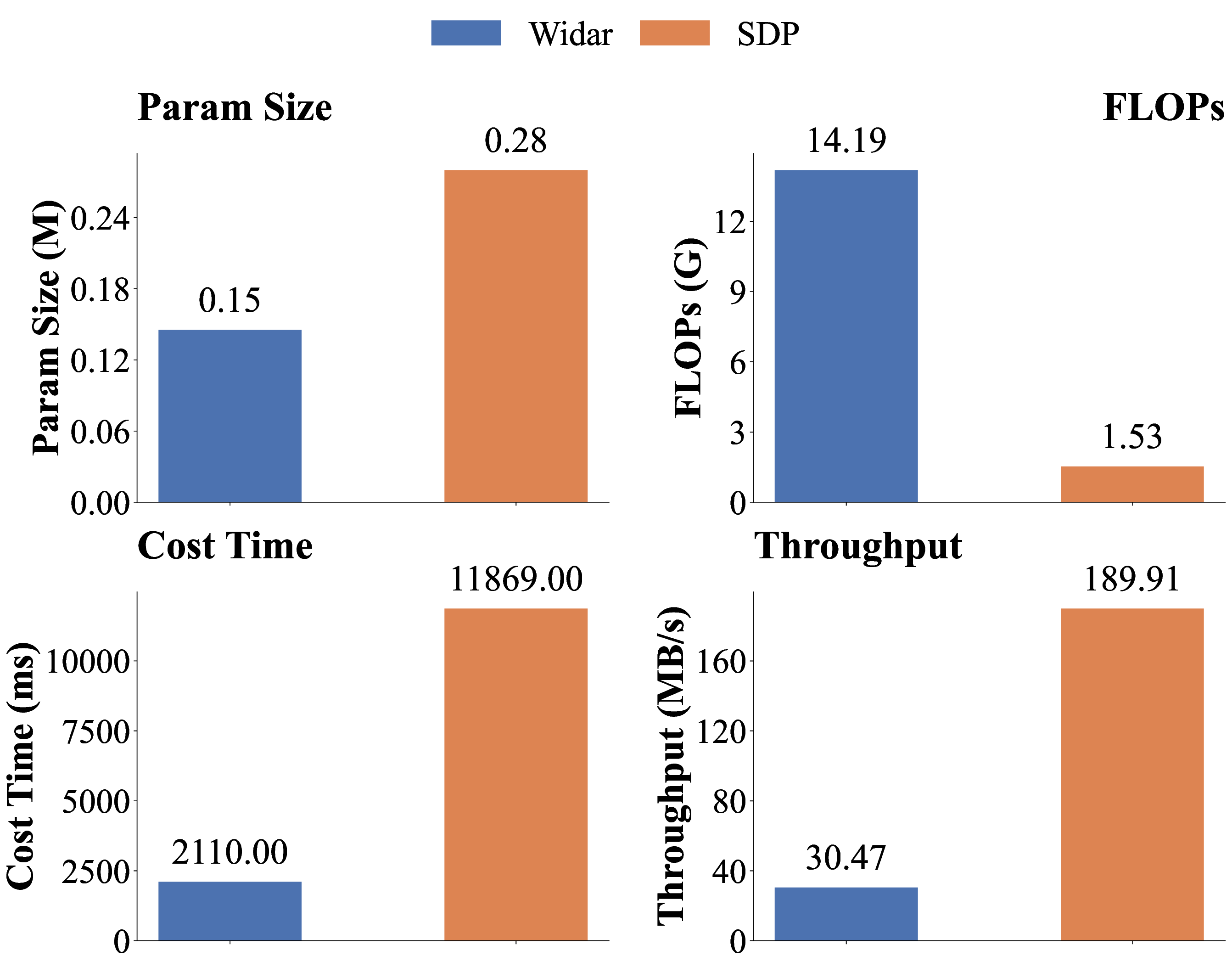}  
    \caption{Analysis of Model Complexity and Efficiency.}
    \label{fig:res_performance}
\end{figure}

\begin{figure}[htbp]
    \centering
    \captionsetup{justification=raggedright}
    \includegraphics[width=0.8\linewidth]{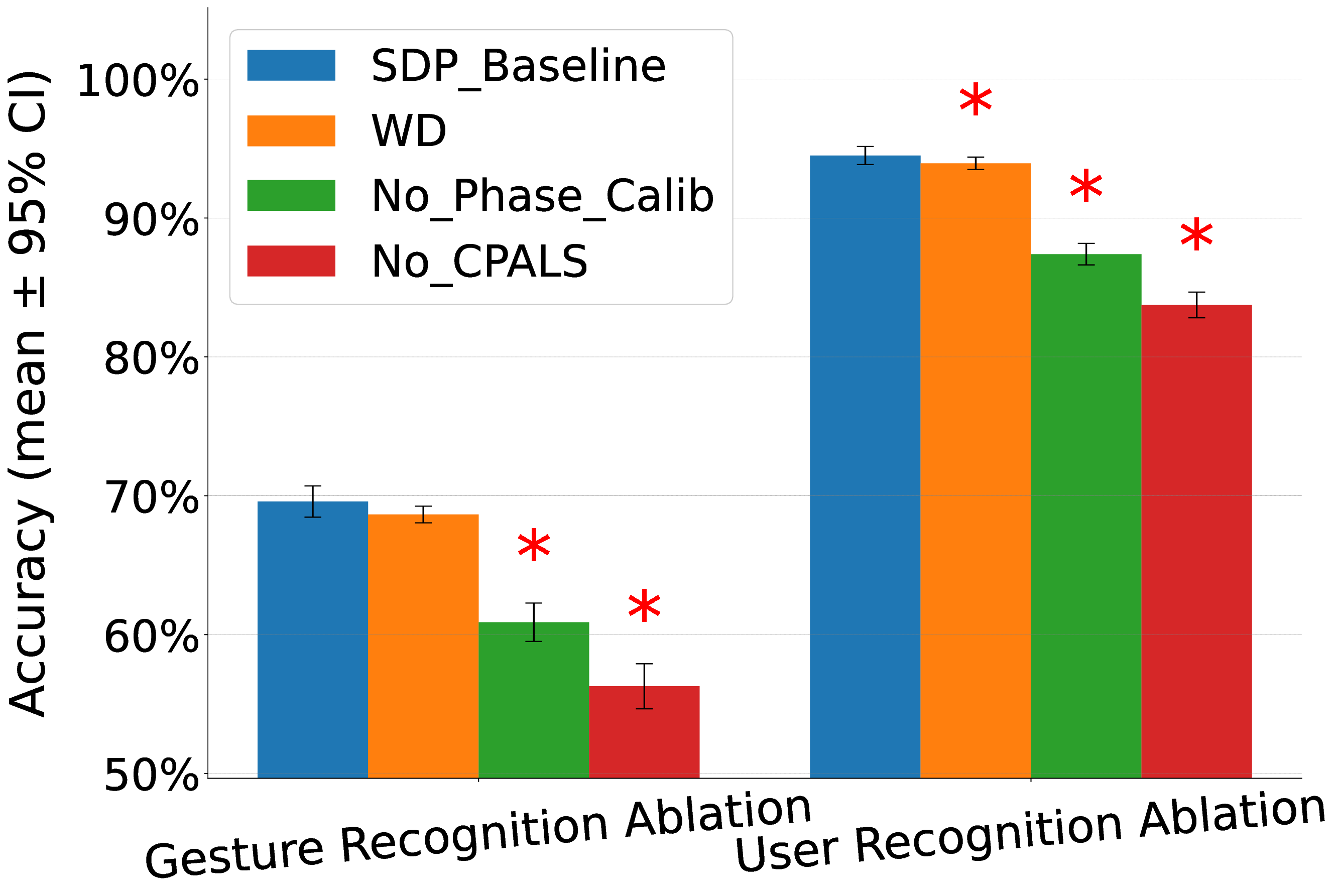}
    \caption{Ablation on Widar and Gait. Removing CP–ALS or phase calibration degrades accuracy; the full SDP model achieves the highest mean accuracy with the smallest CI across seeds. Error bars: 95\% CI; asterisks: paired $t$-test vs.\ full model ($p<0.05$).}
    \label{fig:ablation}
\end{figure}

Efficiency is further analyzed at a matched operating point (Top-1 $\approx$ 69.2\%), with latency measured by CUDA events (batch=1, post warm-up). According to Fig.~\ref{fig:res_performance}, while the richer information in CSI data results in a longer inference time compared to the baseline Widar model, our SDP model demonstrates a clear advantage in FLOPs and offers greater throughput.

Finally, the ablation study in Fig.~\ref{fig:ablation} confirms that CP–ALS pooling and phase calibration are the most critical components, as their removal causes the largest performance drop. The complete SDP configuration offers the most favorable balance of accuracy, stability, and reproducibility, establishing a strong baseline for future work.

\section{Conclusion}
\label{section5}
This paper presented SDP, a protocol and benchmark framework that standardizes multi-task wireless sensing. SDP introduces a flexible data schema and a canonical tensor projection method to unify heterogeneous sensing measurements. Our baseline model, built upon this protocol, was experimentally validated across detection, recognition, and vital-sign estimation tasks. The results demonstrate that our unified pipeline significantly enhances training stability and inference efficiency while maintaining competitive performance, thereby providing a reproducible foundation for fair comparative research.


%

\ifCLASSOPTIONcaptionsoff
  \newpage
\fi

{
\small
\bibliographystyle{IEEEtran}
\bibliography{bibtex/bib/IEEEabrv,bibtex/bib/IEEEexample}
}
\end{document}